\documentclass[aps,
reprint,
amsmath, 
amssymb,
nofootinbib,
raggedbottom,
superscriptaddress]{revtex4-2}
\usepackage{graphicx} 
\usepackage{preamble}
\usepackage{color}
\usepackage[dvipsnames]{xcolor}
\usepackage{cancel}
\usepackage{orcidlink}
\usepackage{soul}

\usepackage{mathtools}

\usepackage{bm}

\begin{document}

\title{Generalized Yang-Mills theory: Interpolating between SDYM and YM}

\author{Tolga Domurcukg\"ul \orcidlink{0000-0002-4852-3818}}
\affiliation{Department of Physics, North Carolina State University, Raleigh, NC 27607, USA \looseness=-1}
\author{Hao Geng \orcidlink{0000-0001-9251-5762}}
\affiliation{Gravity, Spacetime, and Particle Physics (GRASP) Initiative, Harvard University, 17 Oxford St., Cambridge, MA, 02138, USA. \looseness=-1} 
\author{Mendel Nguyen \orcidlink{0000-0002-7976-426X}}
\affiliation{Department of Mathematical Sciences, Durham University, Durham DH1 3LP, UK \looseness=-1}
\author{Mithat \"{U}nsal \orcidlink{0000-0002-4875-9978}}
\affiliation{Department of Physics, North Carolina State University, Raleigh, NC 27607, USA \looseness=-1}

\begin{abstract} 

We construct a generalized Yang-Mills (YM) theory with two real couplings, interpolating continuously between the Self-Dual Yang-Mills (SDYM) limit (also called Chalmers-Siegel theory) and physical Yang-Mills theory. The kinetic coupling $\epsilon$ controls local fluctuations and anti-instanton weight,  while the topological coupling $g$ controls the instanton weight. 
Both couplings are asymptotically free. We derive an exact all-order relation between the beta functions of the two couplings, revealing a Renormalization Group invariant,  a new dimensionless expansion parameter 
$\Lambda_\epsilon / \Lambda_g$ into the study of YM theory. 
In the SDYM limit, the vacuum is  populated by a finite density of topological defects, yet local correlators decay algebraically, consistent with a non-unitary conformal field theory. We confirm this mechanism via compactification on arbitrary size $\mathbb{R}^3 \times S^1$, where the vacuum maps to a non-interacting ideal gas of monopole-instantons. 
As the kinetic coupling is turned on, a mass gap and confinement scale emerge. 

\end{abstract}

\maketitle
\newpage

\subsection{Introduction}

Standard Yang-Mills theory (setting $\theta$ to zero) is a one-parameter theory, and hence, by asymptotic freedom, a one-scale theory. There are no other dimensionless parameters that we can add, and therefore, no known expansion parameters for finite rank gauge groups,    rendering the theory analytically  intractable. 

However, the field strength in YM theory transforms in a reducible representation of the Lorentz group, $(1,0) \oplus (0,1)$. Hence, the field strength can be viewed as a chiral pair, much like a Dirac fermion can be viewed as a pair of chiral fermions. In this work, we begin by projecting out the anti-self-dual sector (negative helicity gluons and anti-instantons). The resulting theory is Self-Dual Yang-Mills (SDYM) \cite{Chalmers:1996rq}, which has been studied extensively in the context of perturbative QCD amplitudes \cite{Bardeen:1995gk,  Witten:2003nn, Bern:1993qk, Mahlon:1993si, Dixon:2024tsb, 
Krasnov:2016emc,  Cangemi:1996rx, Monteiro:2022nqt,Chattopadhyay:2020oxe, Chattopadhyay:2021udc,Bittleston:2020hfv,Bu:2023vjt}. However, there exists virtually no literature on the non-perturbative aspects of SDYM.
SDYM is a very special theory. Classically, its S-matrix is trivial and it possesses an infinite-dimensional symmetry 
\cite{Yang:1977zf, Dolan:1983bp, Nair:1988bq,Ward:1977ta}.  Quantum mechanically, it is saturated at one-loop order, which  Bardeen interpreted as integrability anomaly \cite{Bardeen:1995gk,Monteiro:2022nqt}.  The theory is essentially a non-unitary, logarithmic conformal field theory (CFT)~\cite{Frenkel:2006fy, Frenkel:2008vz,Losev:2017qrj}. Regardless, it is a useful tool for amplitude studies.

In this work, we reintroduce the anti-self-dual sector in a continuous manner, constructing a two-parameter version of Yang-Mills theory that interpolates between SDYM and pure YM.

\subsection{From pure YM to SDYM}
We start with the Euclidean formulation of YM theory, with action: 
\begin{equation}
    S_{\rm YM} = \frac{1}{2g^2} \int d^4x \Tr F^2
\end{equation}
(Throughout, we abbreviate tensor contractions like $\xi_{\mu\nu} \eta_{\mu\nu}$ as $\xi \eta$.) 
At this stage, $g$ is dimensionless physical coupling constant.  Originally,  Chalmers and Siegel 
\cite{Chalmers:1996rq}  obtained SDYM theory in two steps, first by chiral (positive-helicity) truncation ${\cal N}=4$ Super Yang-Mills, and then by
elimination of all lower spin fields. 
For the twistor space perspective, see \cite{Witten:2003nn,Boels:2006ir}. 
Below, we derive the SDYM theory by  projecting out anti-self dual sector from pure YM,  by inserting the  constraint $\delta(F^-)$ into path integration that enforces anti-self-dual fields to vanish:
\begin{align}
 Z_{\rm SD}=\int \mathcal{D}A \;  \exp[-   S_{\rm YM}] \; \delta(F^-)  \,.
\end{align}
This naively implies the removal of, in Euclidean space, all anti-instantons, and in Minkowski, all negative helicity gluons. 
In fact, analytic continuation tells us that both  perturbative and non-perturbative anti-selfdual sectors are removed. 
Using the fact that $\Tr(F^2) = 2\,\Tr (F^-)^2 + \Tr(F \widetilde{F})$, and a Lagrange multiplier field to  impose the constraint, we have
\begin{multline}
    Z_{\text{SD}} = \int \mathcal{D}A  \int \mathcal{D}B \, \exp\left( i \int \Tr B F^-  \right)  \\
    \times \exp\left( - \frac{1}{g^2} \int \Tr(F^-)^2 - \frac{1}{2g^2} \int \Tr F \widetilde{F}  \right) \,.
\end{multline}
Now, the interesting thing is, we can absorb  $\Tr (F^-)^2$ term by  
$\Tr(B F^-)$ term  by a field redefinition.   We choose $B= -\tilde B$, anti-selfdual, hence, we can write $B F^-= B F $. 
Then, the chiral half of partition function becomes:
  \begin{align}
Z_{\text{SD}} = \int \mathcal{D}A   \mathcal{D}B \,  \exp \left( i \int \, \Tr(B F^-)  - \frac{1}{2g^2} \int \Tr(F \widetilde{F}) \right)\,.
\label{SDYM}
\end{align} 
 The Lagrangian  of the SDYM theory, for perturbative purposes,  can be written as:  
 $ -i  \Tr(B F)$ as is used in literature. 
 However, the full action of the self-dual theory \eqref{SDYM} must include the topological term $\Tr(F \widetilde{F})$. The reason is, by our construction,  
SD-theory is a chiral half of YM, and  does not have a parity symmetry, therefore  $\Tr(F \widetilde{F})$ term is not forbidden (unlike pure YM at $\theta=0$).  In fact, even in a case it is not included in the action, it must be generated perturbatively in the full theory \cite{Losev:2017qrj}.  We will see that it does.

In other words,   $\Tr(F \widetilde{F})$ term, being  topological,  does not alter perturbation theory, however,  perturbation theory alters  $\Tr(F \widetilde{F})$. This is a phenomenon that does not occur in pure YM, and 
it will be important in our full construction. 

In particular, the parameter $g^2$  should now be 
 interpreted as a topological coupling, since it appears only in topological term. 
 In fact, there is an exact-1 loop running of  $g^2$  if and only if  the theory is considered in an instanton background, as shown in an underappreciated important work of Losev et.al.\cite{Losev:2017qrj}. This is also confirmed 
 in a recent work of Bittleston and Costello \cite{Bittleston:2025jmk}, 
 using  twistor theory and an index theorem. 
 We also   
 have an independent confirmation of these results by calculating path integral directly in an instanton background, where we map the problem to NSVZ-type computation \cite{Novikov:1983uc}.

It is observed in \cite{Chalmers:1996rq}  that 
 if we deform the SD-action with 
\begin{equation}
    \Delta S=  \int  \frac{g^2}{2}  \Tr B^2 
    \label{special}
\end{equation}
the theory reduces back to pure YM theory.  This is natural, because integrating out 
$B$, it transmutes to  $F^-$  (negative helicity gluon), whereas the already existing $A$ field in the theory is the  positive helicity gluon.  In other words, the deformation \eqref{special}  reintroduces the gluon exactly symmetrically.

\subsection{Generalized Yang-Mills}
In the above section,  the relation between SDYM and YM is a digital one, one is either in one or the other as in a binary. We want to generate an analog version of this, where 
we  introduce a  deformation with a  tunable coupling $ \epsilon$. 
 This deformation, in a similar spirit, is introduced by Witten  to reincorporate negative helicity gluons back into the self-dual theory \cite{Witten:2003nn}.  In our work, we introduce it to reincorporate anti-instantons into the theory. Of course, the two are not unrelated. We are both reincorporating in effect the full anti-selfdual 
sector of the theory by doing so. However, different from earlier works, we take this deformation as a stand-alone master theory interpolating between SDYM and pure YM, and study its remarkable structure associated with its  $\beta$-functions for  $\epsilon$ (kinetic coupling) and $g$ (topological coupling), and non-perturbative aspects of the theory.  
Most importantly, this construction introduce a new dimensionless expansion parameter into the study of YM theory. 
We refer to the resulting two-coupling theory as generalized Yang-Mills (GYM) theory and examine its perturbative and non-perturbative properties below.

 
The partition function of the interpolating theory is:
\begin{align}
    Z_{\rm GYM} = \int \mathcal{D}A   \exp\left( - \frac{1}{\epsilon^2} \int \Tr(F^-)^2 - \frac{1}{2g^2} \int \Tr F \widetilde{F} \right) .
    \label{GYM}
\end{align}
The resulting theory does not have parity symmetry, the two helicity sectors are treated asymmetrically except for $\epsilon=g$. It  interpolates between SDYM and YM:
  \begin{align}
  &    \lim_{\epsilon \rightarrow 0} Z_{\rm GYM}  = Z_{\rm SDYM}, \;\; 
        \lim_{ \frac{\epsilon}{g} \rightarrow 1} Z_{\rm GYM }  = Z_{\rm YM} \;.
  \end{align} 
Recall that $\epsilon=0$ is  non-unitary, and indeed, our GYM theory is non-unitary as well. However, we can express the action of the theory \eqref{GYM} as 
 $\frac{1}{2\epsilon^2 } \Tr(F^2) + (\rm topological \; term)$. Thus, the difference between non-unitary theory and pure YM is a topological (total derivative) term.   Therefore, the GYM theory has the same perturbative structure as ordinary YM, but, non-perturbatively, it is different, as we explain now. 
 

In our construction,  $\epsilon \neq g$  generates  a disparity between instantons and anti-instantons.  Calculating instanton and anti-instanton actions in GYM, we find 
\begin{align}
S_I &= \frac{8 \pi^2}{g^2},  \cr
S_A &= \left( \frac{2}{\epsilon^2}  -   \frac{1}{g^2} \right)  8 \pi^2 
\label{Ins-ac}
\end{align}
  At $\epsilon=0$, anti-instanton action diverges, it decouples from the dynamics. And we land on SDYM theory.  At $\frac{\epsilon^2}{g^2}=1$,  $S_A = S_I$, and the symmetry between SD and anti-selfdual sector is restored, i.e., we we land on pure YM theory with an exact parity symmetry. For $\epsilon > \sqrt2 g$, anti-instanton action becomes negative, and this is the regime of a instability.  Therefore, sensible theories live in the interval  
$0\leq  \epsilon \leq \sqrt 2 g$. 

 
  \subsection{$\beta$ functions for  kinetic and topological couplings}
The first remarkable feature of GYM is the structure of its beta functions. 
As already mentioned, $\epsilon$ is now the standard perturbative coupling,  while $g$ is a topological coupling. Both couplings run as a function of the renormalization group scale. Since 
$g$ is topological, it does not enter into standard perturbation theory, and the beta function of $\epsilon$ is given by: 
\begin{align}
  \beta_\epsilon(g, \epsilon) &=  \frac{1}{(4 \pi)^2}\left( - \beta_0 \epsilon^3 - \beta_1  \epsilon^5 - O( \epsilon^7) \right) \,.
\end{align} 
Furthermore, the running of $g$ is more interesting. Even for $\epsilon=0$, $g$ runs in an instanton background \cite{Losev:2017qrj, Bittleston:2025jmk}. This running is 1-loop exact, and is saturated by zero mode counting and 1-loop fluctuation determinant. There are no further quantum fluctuations around instantons at $\epsilon=0$.   When $\epsilon>0$ is turned on, 
the crucial point to realize is that the structures of cubic and quartic fluctuations are dictated only by $\epsilon$. Expanding the action $S[A]$ around the classical instanton solution $A^{\text{cl}}$:
\begin{equation}
    A = A^{\text{cl}} + a
\end{equation}
and moving to a perturbative normalization of the action, 
we have:
\begin{align}
S[A] = S[A^{\text{cl}}] + S_2[a] + \epsilon S_3[a] + \epsilon^2 S_4[a] + \dots
\end{align}
where  $S[A^{\text{cl}}]$ are the  classical actions \eqref{Ins-ac},  $S_2[a]$ is 
 (coupling independent) fluctuation determinant,  $S_{n}[a] \equiv S_n[A^{\text{cl}}, a]$.
 Cubic and quartic terms control the perturbative  fluctuations around instantons, starting at two-loop order. 
The one-instanton effective action at two-loop orders takes the form (tweaking results of \cite{tHooft:1976snw,Novikov:1983uc} according to the above rules).
 \begin{align}
    \Gamma_1 = \frac{8\pi^2}{g^2} - \beta_0 \ln(\mu \rho) - \beta_1 \frac{\epsilon^2}{16\pi^2} \ln(\mu \rho) \,.
 \end{align}
The one-instanton effective action, $\Gamma_1$, must be independent of the arbitrary renormalization scale $\mu$,
$ \mu \frac{d\Gamma_1}{d\mu} = 0$.  This results in $\beta$ function of $g$.
\begin{align}
 \beta_g(g, \epsilon) &=  \frac{1}{(4 \pi)^2}\left( - \beta_0 g^3 - \beta_1 g^3 \epsilon^2 - O( g^3 \epsilon^4) \right)  \,.
\end{align} 
Note that in the running of the topological coupling $g$, the $\epsilon$ correction  terms play the same role as the one-loop anomalous dimension $\gamma$ in the NSVZ formula—it's the perturbative correction to the instanton-derived beta function \cite{Novikov:1983uc}.


It is also remarkable that 
\begin{align}
 \frac{\beta_g }{g^3} &=  \frac{\beta_\epsilon}{\epsilon^3}  
 = f(\epsilon)  
 \label{per-top}
\end{align}
where $f(\epsilon) = \frac{1}{(4 \pi)^2} (- \beta_0  - \beta_1  \epsilon^2 - \beta_2 \epsilon^4 + \cdots )$.
This is an all-orders relation in our GYM theory. 
This essentially means perturbation theory in instanton background yields the exact values 
$\beta_0, \beta_1,..$  that is known from standard  perturbative calculations, as it must. 

\noindent
{\bf Proof:} Since an all-order relation is a significant result, we want to prove it rigorously. 
Our logic rests on the perturbative non-renormalization of the topological term. The non-renormalization  holds in any renormalization scheme that respects the topological quantum number (such as $\overline{\rm MS}$, background field method, 
perhaps in lattice too \cite{Chen:2024ddr}).
First, rewrite  the Lagrangian density as:
\begin{align}
    \mathcal{L} = \frac{1}{2\epsilon^2} \Tr(F^2) + \left( \frac{1}{2g^2} - \frac{1}{2\epsilon^2} \right) \Tr(F \widetilde{F})\,.
   \label{GYM-new}
\end{align}
In perturbation theory, the term $\Tr(F \widetilde{F})$ is a total derivative, and its coefficient does not receive radiative corrections at any order in perturbation theory. Consequently, the difference between the two couplings is RG invariant:  
\begin{align}
\frac{d}{d \ln \mu} \left( \frac{1}{g^2} - \frac{1}{\epsilon^2} \right) = 0\,.
\end{align}
 Performing the differentiation  yields the exact relation we found in \eqref{per-top}.

\begin{figure}[tbp] 
\begin{center}
\includegraphics[width=0.25\textwidth]{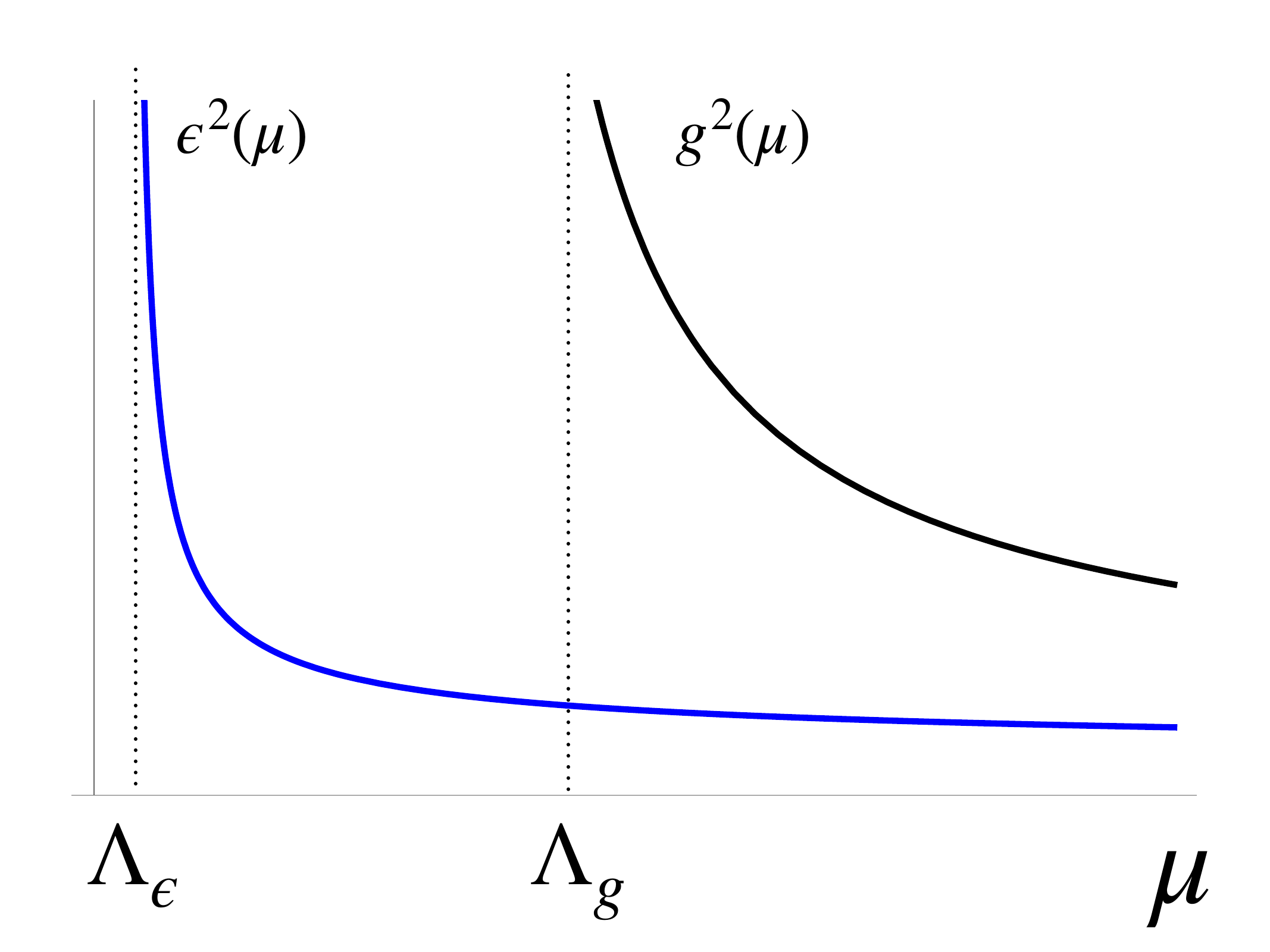} 
\end{center}
\vspace{-0.5cm}
\caption{The running of kinematic and topological couplings, and associated strong scales. 
  }
\label{fig:run}
\end{figure}
Of course, these runnings imply the existence of two-strong scales in GYM theory. One is the standard perturbative strong scale $\Lambda_{\epsilon}$, and the other is the strong scale for topological coupling $\Lambda_{g}$: At 1-loop level, 
\begin{align}
    \Lambda_g = \mu \exp\left[ - \frac{8 \pi^2}{g^2(\mu) \beta_0} \right], \quad   \Lambda_\epsilon = \mu \exp\left[ - \frac{8 \pi^2}{\epsilon^2(\mu) \beta_0} \right] .
\end{align}
The crucial point is that in the $\epsilon \rightarrow 0$ limit, $\Lambda_{g}$ remains robust, actually it is 1-loop exact, while 
$\Lambda_{\epsilon} \rightarrow 0$.  This is reflecting the behavior of the $\beta$ function in SDYM theory.  For $\epsilon=g$, the theory has an extra parity symmetry, and the two couplings are identical. The theory reduces to pure YM theory, with  $\Lambda_{\epsilon} = \Lambda_{g}$, a one scale theory.

\subsection{Scale anomaly}
As we described, the theory develops two strong scales for GYM theory. Hence, the scaling current, which is conserved classically, is no longer conserved in the quantum theory. The trace anomaly for GYM theory is sourced by the running of both kinetic and topological couplings:
\begin{align}
 (T^\mu_\mu)_{\rm GYM} 
& = 2 \frac{\beta_\epsilon}{\epsilon^3}  \Tr(F^-)^2  + \frac{\beta_g}{g^3}  \Tr F \widetilde{F}  \\
&= f(\epsilon)  \Tr F^2 \,.
\label{scalean}
\end{align}
as per \eqref{GYM} and \eqref{per-top}.


This result is consistent with both SDYM theory $\epsilon=0$, and pure YM theory, with a crucial difference between the two limits. 
Note that despite the fact that $\beta_\epsilon \rightarrow 0$ and $\Lambda_\epsilon \rightarrow 0$ in the $\epsilon \rightarrow 0$ limit, the effect of the perturbative term does not die off in this limit. What matters is the ratio $\beta_\epsilon/\epsilon^3$, which has a finite limit saturated at 1-loop, given by $\frac{1}{(4 \pi)^2} (- \beta_0)$. Remarkably, this combines with the running of the topological coupling to yield the exact 1-loop result in SDYM theory:
\begin{align}
 (T^\mu_\mu)_{\rm SD} &= - \frac{1}{(4 \pi)^2} \beta_0  \Tr F^2  \qquad \text{(1-loop exact)}.
\label{scalean-SDYM}
\end{align}
In the self-dual limit,  the effective action for scale-anomaly \cite{Riegert:1984kt, Monteiro:2022nqt,Nair:1990aa} generates the non-vanishing rational one-loop amplitudes which constitute the first non-trivial quantum corrections to the self-dual sector \cite{Monteiro:2022nqt,Cangemi:1996rx}. 
Therefore, we interpret Bardeen's integrability anomaly \cite{Bardeen:1995gk} (which is 1-loop exact) 
as $\epsilon \rightarrow 0$  limit of \eqref{scalean}. It is 
continuously connected to the standard scale anomaly   in pure YM theory (which receives contributions to all orders). 
We also remark that in the context of twistor string theory one often seeks to cancel this anomaly (via an axion field) to preserve quantum integrability \cite{Costello:2021bah, Costello:2022wso}. 
In the context of SDYM theory in 4d, it is a physical property of the theory.


\subsection{Is there a mass gap in SDYM on  $\mathbb R^3 \times S^1$ and $\mathbb R^4$?}
We would like to know non-perturbative properties of  the SDYM theory  $(\epsilon=0)$. Do the properties  such as the mass gap and  confinement of pure YM theory survive in the SDYM theory on ${\mathbb R}^4$ or  ${\mathbb R}^3 \times S^1$? We will show that the answer is, no, despite the fact that the vacuum is filled with a finite density of monopole-instantons.  We will show this explicitly on $\mathbb R^3 \times S^1$ (for any value of $S^1$ size) and reliably extrapolate the answer to $\mathbb R^4$.  

On $\mathbb R^3 \times S^1$, the first question to answer is whether a gauge holonomy potential 
gets generated in SDYM.
The answer is no, and simple to see. 
Using the  action \eqref{GYM} in the background $\Phi = LA_4= {\rm diag}(v_1, \ldots, v_N)$,  
  one can compute the holonomy  potential as in \cite{Gross:1980br,Kovtun:2007py} for finite $\epsilon$. Under the canonical normalization for the kinetic term, the gauge holonomy is  
  $ \Omega= \exp [ i \epsilon \Phi]$, and the effective  1-loop potential takes the form (in theory with $N_f$ adjoint representation fermions with periodic boundary conditions)
\begin{align}
{\mathcal V}(\Omega) = \frac{2 (N_f - 1)}{\pi^2 L^4} \sum_{n=1}^{\infty} \frac{1}{n^4} \left| \text{tr}(e^{  i  \epsilon n \Phi} )\right|^2
\end{align}
Hence, regardless of whether the potential is center-stabilizing $(N_f >1)$, or center-breaking  $(N_f =0)$, or marginal  $(N_f =1)$ for finite $\epsilon$, the potential in  the SDYM or SDQCD limit ($\epsilon \rightarrow 0$) is always flat, to all orders in perturbation theory. Hence, 
the potential is zero (up to an overall constant):
\begin{align}
\mathcal{V}(\Omega)= 0
\end{align}
Due to the vanishing of the holonomy potential, the gauge structure is generically reduced to $U(1)^{N-1}$, there are $N$ types of monopole-instantons, and no anti-monopoles \cite{Kraan:1998sn, Lee:1997vp}.    In the YM theory, we can use abelian duality to write the Maxwell term 
as a kinetic term for the gauge holonomy and dual photon scalars, $z=  \phi - i \sigma$,  where $z$ is $N-1$ component  field \cite{Unsal:2008ch, Unsal:2007jx,Unsal:2007vu}, similar to Polyakov model \cite{Polyakov:1975rs}.
 In Yang-Mills theory,  at the classical level,  monopole operator is holomorphic, and anti-monopole is anti-holomorphic.   However, this fact is usually ignored  because $\phi$ acquires a perturbative mass quantum mechanically (except for supersymmetric theories).

In SDYM, only self-dual monopoles are present. Their effect in the effective action is captured by holomorphic operators 
\begin{align}
&{\cal M}_j (x) =  \exp\left[ -\frac{1}{\epsilon} \alpha_j \cdot z(x) \right], \qquad  \alpha_j \in \Delta_{\rm aff} 
\end{align} 
The proliferation of monopoles generates a holomorphic potential, obeying the constraint  $\prod_j  {\cal M}_j (x)=e^{-S_I}$. The holomorphic potential has $N$-critical points, at which  
$ e^{ -\frac{1}{\epsilon} \alpha_j \cdot z_{{\rm cr}, k}(x)} \equiv \zeta_{j,k}= e^{-S_I/N + i 2\pi k/N}, \;  k=0, \ldots, N-1$, which are the remnants of $N$-branch structure in the SDYM limit. From now on, we restrict to a fixed $k$. The density of the monopoles is 
 controlled by $ \zeta = (1/L^3) \exp\left[{-{8 \pi^2}/{g^2N}}\right]$. We can write an effective action around an arbitrary critical point as:
\begin{align}
S = \int d^3x \left[ |\nabla z|^2 -  \zeta \sum_{\alpha_i \in \Delta_{\rm aff} } \exp\left[ -\frac{1}{\epsilon} \alpha_i \cdot z(x)\right] \right]
\end{align}
where the potential is holomorphic, since it is based on self-dual configurations. 
We define the interacting partition function $Z$ relative to the free Gaussian partition function $Z_0$ (defined by the kinetic term $|\nabla z|^2$). We expand the exponential potential term in powers of the fugacity $\zeta$:
\begin{align}
\frac{Z}{Z_0} = \left\langle  \prod_{\alpha_i \in \Delta_{\rm aff}} \exp  \int d^3x \; \zeta   \exp\left[ -\frac{1}{\epsilon} \alpha_i \cdot z(x)\right]      \right\rangle_0\,.
\end{align}

Expanding exponentials creates a sum over  all monopoles ${\cal M}_i (x)$.  The crucial point is to realize that the two-point function of the holomorphic field is zero, $\langle z^a(x) z^b(y) \rangle_0 = 0$.  For two BPS monopoles, $\langle e^{-\alpha_iz(x)} e^{ - \alpha_j z(y)} \rangle_0 = 1$, the $1/r$ interaction due to $\sigma$ exchange cancels exactly the one due to $\phi$ exchange.
This implies that all interaction terms are trivial, monopoles  do not see each other. Summing over all monopoles, we find: 
\begin{align}
Z_{\rm SD} = Z_0 \cdot e^{N \zeta V} \qquad   ( \rm ideal \; gas)\,. 
\end{align} 
The Coulomb gas representation of SDYM theory corresponds  to a non-interacting ideal gas!
 The correlation length is infinite  because the system remains free and massless, despite the vacuum being filled with monopoles.  Summing over all critical points of the holomorphic potential, we find (restoring $\theta$ angle): 
\begin{align}
Z_{\rm SD}(\theta) = Z_0 \sum_{k} \exp \left[ N V \frac{1}{L^3}  e^{-\frac{S_I}{N} + i \frac{\theta + 2 \pi k}{N}} \right]
\end{align} 
which is the contribution of self-dual configurations to  the vacuum energy density in semi-classical regime of 
YM theory \cite{Unsal:2008ch}. 

This conclusion holds for any size of $\mathbb R^3 \times S^1$.
This is because the monopole effective field theory is valid provided $LN  \Lambda_\epsilon \lesssim 1$. 
Since in SDYM, we have $\Lambda_\epsilon=0$, due to $\epsilon=0$, the condition $LN  \Lambda_\epsilon \lesssim 1$ is trivially satisfied. 
Hence, the result holds for any size of $\mathbb R^3 \times S^1$. 
This is a non-perturbative confirmation that SDYM theory on $\mathbb R^4$ remains a (non-unitary) CFT \cite{Losev:2017qrj,Costello:2022wso}. 
 
Once an $\epsilon$ is turned on,  GYM theory becomes trickier due to the fact that the finite 
$\epsilon \ll 1$ theory prefers spontaneous breaking of the center-symmetry for sufficiently small $\mathbb R^3 \times S^1$, where any non-trivial gauge holonomy background becomes unstable. However, the story in SDQCD(adj) is conceptually very clean and elegant. With $\epsilon \ll 1$, non-trivial holonomy background is always stable \cite{Kovtun:2007py}.  There are both  monopoles, and anti-monopoles, unsuppressed/suppressed respectively, and  the bion amplitudes are also suppressed. Their densities in vacuum are:
\begin{align}
&{\cal M}_i (x) \sim  \exp\left[-\frac{S_I}{N} \right] =\exp\left[ -\frac{8 \pi^2}{g^2N} \right]  \,, \cr 
& \overline{\cal M}_i (x)  \sim  \exp\left[-\frac{S_{AI}}{N}\right] = \exp\left[- \left( \frac{2}{\epsilon^2} -\frac{1}{g^2} \right)  \frac{8 \pi^2}{N} \right]\,, \cr 
&{\cal B}_{i, i+1} = [{\cal M}_i \overline{\cal M}_{i+1} ]  \sim \exp\left[- \left( \frac{2}{\epsilon^2}\right)  \frac{8 \pi^2}{N} \right]\,.
\end{align}
In the $\epsilon \rightarrow 0$ limit, only ${\cal M}_i (x)$ survives. They do not lead to a mass gap and confinement (both because of holomorphy and fermionic zero modes). At $\epsilon  \ll 1$, both players are in the game, but with a disparity.  But monopoles and anti-monopoles won't generate a mass gap by themselves, due to their fermi zero modes \cite{Affleck:1982as,Unsal:2007jx, Unsal:2007vu}. At second order in semi-classics, 
magnetic bions (two-clusters of monopoles without fermi zero modes) do generate a mass gap, as well as confinement. 
In the generalized QCD(adj), the mass gap  is given by: 
\begin{align}
    m^2_{\epsilon} = \frac{1}{L^2}  \exp\left[-\frac{S_I}{N} -\frac{S_{AI}}{N} \right] 
    = \frac{1}{L^2} \exp\left[-\frac{16 \pi^2}{\epsilon^2N} \right]
\end{align}
which is a quite curious result. Mass gap (square) is proportional to the product of density of monopoles and density of 
anti-monopoles! This tends to zero  in the  $\epsilon\rightarrow0$ limit where anti-monopole density vanish, and magnetic 
bion formation is impossible. 
For generic $\epsilon < g$, it should be noted that this mass gap is exponentially smaller than the mass gap in standard 
QCD(adj) on  $\mathbb R^3 \times S^1$.  

It remains to be seen if we can use this framework on the mass gap and confinement problem in Yang-Mills theory on $\mathbb R^4$, by perturbing  the  conformal theory at $\epsilon=0$, and using techniques of Refs.~\cite{Nguyen:2023rww, Nguyen:2025voy}. In particular, in the generalized theory, the classical interactions take exactly the same form as in ordinary Yang--Mills theory---zero between like-chirality instantons and dipole--dipole--like for opposite-chirality instantons---except that the strength of the interactions are given by the factor $1/\epsilon^2$ rather than $1/g^2$, while the density of instantons is controlled by $1/g^2$.

\subsection{Outlook}
We have constructed a generalized Yang-Mills theory establishing a rigorous bridge between the conformal and the full confining dynamics. 
Our central result, the exact relation $\beta_g/g^3 = \beta_\epsilon/\epsilon^3$, reveals an RG invariant separating the confinement scale from the topological charge density of self-dual configurations. 
This framework resolves the characterization of the SDYM limit: the vacuum hosts a finite density of self-dual topological defects, yet the fact that they are exactly non-interacting, does not harm  the structure of  algebraic correlators---the hallmark of a non-unitary CFT.  
Turning on the kinetic coupling disturbs this balance; hence,  this formalism continuously maps the transition from conformality to confinement.   

We are essentially proposing to study Yang-Mills theory not as a perturbation of a free theory, but as a perturbation of a logarithmic CFT, the SDYM theory, for which the vacuum is filled with instantons.



\acknowledgments
We thank Pratik Chattopadhyay, Lewis Cole, Lance Dixon, Yui Hayashi,  Kirill Krasnov, Tin Sulejmanpasic,  Yuya Tanizaki, and Laurence Yaffe for various conversations about this work. 
H.G. is supported by the Gravity, Spacetime, and Particle Physics (GRASP) Initiative from Harvard University.
T.D. and M.\"U. are supported by U.S. Department of Energy, Office of Science, Office of Nuclear Physics, under Award Number DE-FG02-03ER41260. 
M.N. is supported by The Royal Society, University Research Fellowship and STFC consolidated grant number ST/X000591/1. 

\bibliography{QFT-Mithat-2}

\end{document}